\documentclass[11pt]{article}

\usepackage[final]{acl}

\usepackage{times}
\usepackage{latexsym}

\usepackage[T1]{fontenc}
\usepackage[utf8]{inputenc}

\usepackage{microtype}

\IfFileExists{inconsolata.sty}{\usepackage{inconsolata}}{}

\usepackage{graphicx}
\usepackage{booktabs}
\usepackage{multirow}
\usepackage{amsmath}
\usepackage{amssymb}
\usepackage{array}
\newcommand{\codeid}[1]{\nolinkurl{#1}}

\title{How Much Do Modern Embeddings Reveal?\\
Inversion Attacks on Commercial Embedding Models}

\author{Kareem Elaasar \\
  Department of Informatics \\
  University of California, Irvine \\
  \texttt{kelaasar@uci.edu} \\\And
  Cristina Videira Lopes \\
  Department of Informatics \\
  University of California, Irvine \\
  \texttt{lopes@uci.edu} \\}

\begin{document}
\maketitle

\begin{abstract}
Text embedding inversion reconstructs source text from fixed-dimensional vectors, posing a privacy risk to systems that store embeddings instead of raw text. We adapt the vec2text framework \citep{morris2023text} to support \texttt{text-embedding-3-small} and \texttt{mistral-embed} alongside GTR-base. We evaluate 800 samples per category across general instructions, Wikipedia passages, paraphrased instructions, health-related instructions, and short literal code snippets, using a common 60--80-character range and paired original and paraphrased instructions. Under the evaluated vec2text configuration, \texttt{text-embedding-3-small} achieves the highest exact-match rate in all five categories, including 47.4\% on health-related instructions and 24.5\% on general instructions under the evaluation's case-insensitive, whitespace-stripped matching rule. BERTScore F1 exceeds 87 on the 0--100 scale for every prose category and embedder, indicating high contextual similarity to the references. Short code snippets yield lower reconstruction scores. These findings characterize one attack configuration and the selected datasets; they do not establish attack-independent rankings or sensitive-attribute recovery rates. They nevertheless motivate explicit assessment of inversion risk and access controls for stored embeddings.

\end{abstract}

\section{Introduction}

Dense text embeddings are the infrastructure layer of modern NLP deployment. Retrieval-augmented generation (RAG) systems, semantic search engines, and recommendation pipelines all rely on fixed-dimensional vectors computed from raw text---and they routinely store those vectors in persistent databases such as Pinecone, Weaviate, and Chroma. A common implicit assumption in these deployments is that embeddings are \emph{privacy-preserving by construction}: they capture semantic meaning while discarding surface form, making reconstruction of the source text infeasible.

This assumption was directly challenged by \citet{morris2023text}, who showed that a T5-based sequence-to-sequence model---trained to invert embeddings and iteratively corrected by a second model---could recover substantial portions of source text from GTR-base and \texttt{text-embedding-ada-002} embeddings. Their result has significant implications for the real-world trustworthiness of embedding-based systems: if vectors can be inverted, then organizations that store only embeddings (e.g., to comply with data minimization policies) may be inadvertently retaining recoverable text.

Despite this foundational result, two questions motivate further evaluation. First, commercial embedding APIs now include OpenAI's \texttt{text-embedding-3-small} and Mistral's \texttt{mistral-embed}; their reconstruction rates under a common inversion configuration merit comparison. Second, although prior work evaluated inversion across multiple domains, including BEIR datasets and MIMIC-III clinical notes, comparisons across domains can be affected by differences in input length. We complement these evaluations with five categories restricted to the same character-length range and a paired comparison of original and paraphrased instructions.

This paper investigates these questions. We make the following contributions:

\begin{enumerate}
\item \textbf{Support for additional embedding APIs.} We adapt the open-source vec2text framework to support \texttt{text-embedding-3-small} and \texttt{mistral-embed} alongside GTR-base. The implementation adds model-specific embedding dimension selection, API client integration, and checkpoint registration while retaining the original inversion architecture and training procedure.

\item \textbf{Controlled cross-domain evaluation.} We evaluate 800 samples per category across general instructions \citep{zhao2024oneminstructions}, Wikipedia passages \citep{merity2017pointer}, paraphrased instructions, health-related instructions, and short literal code snippets. Inputs are restricted to 60--80 characters, and each paraphrased instruction is paired with its original, supporting comparisons under a common character-length constraint.

\item \textbf{Empirical findings for the evaluated configuration.} With up to five corrector steps, \texttt{text-embedding-3-small} achieves 47.4\% exact match on health-related instructions and 24.5\% on general instructions under the evaluation's normalization rule. Every prose category achieves BERTScore F1 above 87 on the 0--100 scale for all three embedders. Health-related instructions have the highest exact-match rates and short code snippets the lowest among the evaluated categories. These observations are conditional on the attack, datasets, and training runs used here.
\end{enumerate}

This work sits at the intersection of information retrieval and NLP security, providing empirical grounding for updated threat models around the trustworthiness of deployed embedding systems.

\section{Background and Related Work}

\subsection{Text Embedding Inversion}

\citet{morris2023text} introduced vec2text, an iterative approach to reconstructing text from embeddings. Its two-stage architecture consists of an \emph{inverter}, which projects an embedding into pseudo-token representations for T5-base \citep{raffel2020t5}, and a \emph{corrector}, which refines the hypothesis using the target embedding, hypothesis embedding, and hypothesis text. Their strongest reported setting recovered 92\% of 32-token inputs exactly. They also evaluated cross-domain transfer and recovery of personal information from clinical notes. Our results use different evaluation data and inference settings and should not be interpreted as a matched replication of that headline recovery rate.

\citet{morris2024language} studied reconstruction of prompts from language models' next-token probability distributions, including those of Llama-2. This targets a different observable output than the fixed-dimensional text embeddings studied here. Our evaluation uses only vec2text for embedding inversion and does not establish how alternative attack families would rank the same embedders.

\subsection{Privacy and Security of Embeddings}

A growing line of work proposes defenses against inversion, including differentially private embedding mechanisms that add calibrated noise to vectors and adversarial perturbations applied to embeddings before storage. Our work is upstream of these defenses: before evaluating whether a mitigation works, one must characterize the attack surface across the diverse embedders in use today.

Prior analyses of neural representation invertibility in computer vision \citep{dosovitskiy2016inverting,mahendran2015understanding} established that representations preserve more information than expected. Our work establishes an analogous result for modern text embedding APIs, with the added complication that the embedders are black-box commercial systems.

\subsection{Embedding Models}

Table~\ref{tab:embedders} summarizes the three evaluated embedders: two commercial APIs and one historical open-weight baseline. Embeddings used by the inversion pipeline are L2-normalized, including client-side normalization for Mistral. Gemini support was implemented but is not part of the evaluation.

\begin{table}[h]
\centering
\small
\begin{tabular}{lccc}
\toprule
\textbf{Model} & \textbf{Dim} & \textbf{Access} & \textbf{Norm.} \\
\midrule
GTR-base            & 768  & Local & Yes \\
text-emb-3-small    & 1536 & OpenAI API & Yes \\
mistral-embed       & 1024 & Mistral API & Yes \\
\bottomrule
\end{tabular}
\caption{Embedding models evaluated. Vectors are L2-normalized in the evaluation pipeline. GTR-base is loaded locally; API models require HTTP calls per batch.}
\label{tab:embedders}
\end{table}

\textbf{GTR-base} \citep{ni2021sentence} is a T5-based dual encoder producing 768-dimensional embeddings. We use the local Sentence Transformers implementation as a baseline connected to the original vec2text evaluation. Reported inverter throughput is approximately 1,206 samples/second (Appendix~\ref{sec:appendix_training_time}).

\textbf{OpenAI \texttt{text-embedding-3-small}} is accessed through the OpenAI Embeddings API with 1536-dimensional output. We evaluate this output dimensionality without a truncation ablation.

\textbf{Mistral \texttt{mistral-embed}} is accessed through the Mistral API and produces 1024-dimensional embeddings. We do not infer its parameter count or training-data composition from the inversion results.

\subsection{Threat Model}\label{sec:threat_model}

We assume a passive adversary who has obtained one or more embedding vectors $\mathbf{e} = \phi(x)$ produced by a known embedder $\phi$. The adversary has no access to the source text $x$, no access to gradients or internal activations of $\phi$, and no side-channel information. The adversary may call $\phi$ on inputs of their own choosing (a black-box query oracle, as is the case for any commercial embedding API), and may collect a corpus of $(x, \phi(x))$ pairs at training time to fit an inversion model. The adversary's goal is to recover $x$, or a sufficiently close approximation, from $\mathbf{e}$ alone.

This threat model matches the realistic posture of contemporary embedding deployments. Vectors persisted to vector databases (Pinecone, Weaviate, Chroma), cached in memory, transmitted between microservices, or exfiltrated from compromised storage are all instances of the same primitive: a fixed-dimensional vector decoupled from its source text. The adversary need not penetrate the embedding model itself---they need only obtain its output. We do not assume access to the corrector or inverter weights of the target system, and our attack is gradient-free at inference time: each refinement step uses greedy decoding plus a single forward pass through the publicly available embedder.

We deliberately exclude two adversary capabilities: (i) we do not assume access to noise- or perturbation-removal oracles, and (ii) we do not consider active attacks that submit chosen inputs to a victim's RAG pipeline. Both are interesting extensions but lie outside the scope of this work.

\section{Extending Vec2Text to New Embedding APIs}

\subsection{Vec2Text Architecture Overview}

We build directly on the vec2text framework of \citet{morris2023text}, which we summarize briefly for completeness. Vec2text provides two model components: an \texttt{InversionModel} (hypothesizer) and a \texttt{CorrectorEncoderModel}, both implemented as T5-base (220M parameters) encoder-decoders. The inverter maps an embedding $\mathbf{e} \in \mathbb{R}^d$ to an initial text hypothesis by projecting $\mathbf{e}$ through an MLP into 16 pseudo-token slots fed to the T5 encoder. The corrector then refines iteratively, conditioning on the target embedding, the current hypothesis embedding, their difference, and the hypothesis token sequence---allowing it to navigate toward the target in embedding space. Inference is gradient-free; each step uses greedy decoding and halts early when cosine similarity stops improving. We refer readers to \citet{morris2023text} for full architectural details. Our contribution is extending this framework to support new commercial embedding APIs and conducting a systematic cross-domain evaluation.

\subsection{Codebase Adaptations}\label{sec:extensions}

The original vec2text codebase supported only two embedders: GTR-base (local) and \texttt{text-embedding-ada-002} (OpenAI API). The API-specific dimension blocks used a fixed value of 1536. We replace these blocks with model-specific dimension selection. Supporting \texttt{text-embedding-3-small} and \texttt{mistral-embed} required five targeted changes to the codebase; the T5-base backbone, MLP projection scheme, corrector concatenation format, and training loop are all unchanged from \citet{morris2023text}.

\paragraph{1. Dynamic embedding dimension routing (\texttt{inversion.py}).}
The original code hardcoded \texttt{self.embedder\_dim = 1536} for all API-based models. We replaced this with a conditional dispatch based on the \texttt{embedder\_model\_api} argument, branching on each model name to set the embedding dimension (1536 for ada-002 and text-embedding-3-small, 3072 for text-embedding-3-large, 768 for gemini-embedding-001, and 1024 for the mistral-embed family); the full conditional is reproduced in Appendix~\ref{sec:appendix_api}. This value controls the size of the MLP projection layer that maps the embedding into T5's encoder space and must be set correctly for training to proceed.

\paragraph{2. Same fix in \texttt{corrector\_encoder.py}.}
The corrector model contains its own identical hardcoded \texttt{embedder\_dim = 1536} block. The same conditional routing was applied here, since the corrector independently instantiates its projection layers.

\paragraph{3. New API client functions (\texttt{utils.py}).}
Two new embedding functions were added and wired into the existing \texttt{embed\_api()} dispatch function:
\begin{itemize}
  \item \codeid{get_embeddings_mistral()} --- calls \texttt{mistralai.Mistral}, batches in groups of 100, normalizes output, and uses exponential backoff retry (4s$\to$120s, up to 50 attempts) to handle API outages during long training runs.
  \item \texttt{get\_embeddings\_gemini()} --- calls \texttt{google.genai.Client}, batches in groups of 100, normalizes output, uses the \texttt{SEMANTIC\_SIMILARITY} task type with \codeid{output_dimensionality=768}.
\end{itemize}
Both functions follow the same interface as the existing OpenAI embedding function and are dispatched via prefix matching on the API model name; the dispatch code is reproduced in Appendix~\ref{sec:appendix_api}.

\paragraph{4. Checkpoint registry (\texttt{aliases.py}).}
New entries were added pointing to trained checkpoint directories for the evaluated models, allowing them to be loaded via the standard \texttt{load\_corrector()} interface.

\paragraph{5. Public API model list (\texttt{api.py}).}
\texttt{text-embedding-3-small} and \texttt{text-embedding-3-large} were added to the \texttt{SUPPORTED\_MODELS} list exposed by the vec2text public API.

In summary, the only structural changes are (1) dynamic embedding dimension selection and (2) new API client functions. Everything else---T5-base backbone, MLP projection, corrector concatenation scheme, training loop---is identical to \citet{morris2023text}.

\subsection{Training Protocol}

Following \citet{morris2023text}, all models are trained on the MS MARCO Passage Corpus \citep{nguyen2016ms} (\codeid{Tevatron/msmarco-passage-corpus} on Hugging Face), comprising approximately 8.8 million short web passage snippets split 99/1 into train and validation sets. Passages are truncated to a maximum sequence length of 128 tokens using the T5 tokenizer. This is the corpus used in the original vec2text paper. Our comparison uses the training settings reported below; the selected checkpoint steps are listed separately in Appendix~\ref{sec:appendix_checkpoints}. Both inverter and corrector use T5-base (220M parameters) as the backbone. Table~\ref{tab:hyperparams} lists the reported training hyperparameters. The comparison does not isolate architectural or training-data causes of differences in reconstruction performance. All training was conducted on 5$\times$ NVIDIA RTX 6000 Ada Generation GPUs (48 GB VRAM each), CUDA 12.6.

\begin{table}[t]
\centering
\small
\begin{tabular}{lcc}
\toprule
\textbf{Hyperparameter} & \textbf{Inv.} & \textbf{Corr.} \\
\midrule
Base model          & \multicolumn{2}{c}{T5-base}          \\
Batch size/device   & 128           & 16                   \\
Grad.\ accum.\      & 1             & 3                    \\
Eff.\ batch size    & 640           & 240                  \\
Learning rate       & 0.001         & 0.001                \\
Epochs              & 2             & 2                    \\
Warmup steps        & 625           & 20,000               \\
LR scheduler        & const.        & const.               \\
Save steps          & 20,000        & 4,000                \\
Eval steps          & 20,000        & 2,000                \\
Embedder prec.      & \multicolumn{2}{c}{float16}          \\
Model prec.         & \multicolumn{2}{c}{float32}          \\
\bottomrule
\end{tabular}
\caption{Reported training hyperparameters. Corrector effective batch size = 16 $\times$ 5 GPUs $\times$ 3 = 240.}
\label{tab:hyperparams}
\end{table}

\paragraph{Training times.} Reported wall-clock durations vary across models (Table~\ref{tab:training_times}). Inverter training takes approximately 4 hours for GTR-base, 13 hours for \texttt{text-embedding-3-small}, and 15 hours for \texttt{mistral-embed}; corrector training takes approximately 28--48 hours. API access, caching, and model computation can all affect elapsed time. We do not isolate their individual contributions.

\begin{table}[t]
\centering
\small
\begin{tabular}{lcc}
\toprule
\textbf{Model} & \textbf{Inverter} & \textbf{Corrector} \\
\midrule
GTR-base             & $\sim$4h 02m  & $\sim$27h 58m \\
text-emb-3-small     & $\sim$13h 05m & $\sim$48h 28m \\
mistral-embed        & $\sim$15h 01m & $\sim$45h 56m \\
\bottomrule
\end{tabular}
\caption{Wall-clock training times on 5$\times$ NVIDIA RTX 6000 Ada (48 GB). The table reports elapsed time rather than an isolated measurement of API or architectural overhead.}
\label{tab:training_times}
\end{table}

The corrector for each model is initialized from that model's best inverter checkpoint (selected by lowest validation loss on the capped validation set of 500 samples, \texttt{max\_eval\_samples=500}). Before corrector training begins, the trained inverter runs over the entire training set to generate and cache hypothesis predictions to disk---approximately 192 GB for \texttt{text-embedding-3-small}---so that corrector training can proceed without re-running the inverter at each step. The optimizer is AdamW (Hugging Face Trainer default) with cross-entropy language modeling loss (next-token prediction). Training is logged via Weights \& Biases (project: \texttt{emb-inv-4}). Full checkpoint details are in Appendix~\ref{sec:appendix_checkpoints}.

\section{Cross-Domain Case Study}

\subsection{Motivation}

Held-out MS MARCO evaluation provides an in-distribution reference, while prior vec2text work also evaluated multiple other domains. We complement that work by comparing five text categories under a common character-length constraint and pairing original and paraphrased instructions. These controls reduce selected sources of variation without isolating every difference between categories.

\subsection{Dataset Construction}

For correspondence with the released artifacts, \codeid{known_text}, \codeid{unknown_text}, \codeid{medical_text}, and \codeid{paraphrase_text} refer to general instructions, Wikipedia passages, health-related instructions, and paraphrased instructions, respectively. These identifiers do not imply training-set membership.

All datasets are constructed with \texttt{random.seed(42)} and 800 samples per category using the \codeid{prepare_case_study_data.py} script released with this paper. Inputs are restricted to 60--80 characters across categories. This matches the permitted character-length range, not necessarily the token-length distribution or reconstruction difficulty.

\paragraph{General instructions.} We select 800 user instructions from One Million Instructions \citep{zhao2024oneminstructions} (\codeid{wentingzhao/one-million-instructions}, Hugging Face), filtered to 60--80 characters. This category provides a reference for comparison with health-related and paraphrased instructions. It is drawn from a different source than the MS MARCO passages used to train the inverter and is not treated as an in-distribution evaluation.

\paragraph{Wikipedia passages.} We select 800 passages from the WikiText-103 training split \citep{merity2017pointer}, excluding section headers and filtering to 60--80 characters. This category evaluates reconstruction of encyclopedic prose from a source distinct from both the inverter's training corpus and the instruction datasets.

Our evaluation does not include a held-out MS MARCO reference condition. The comparison between general instructions and Wikipedia passages therefore contrasts two evaluation sources distinct from the inverter's training corpus, rather than an in-distribution versus out-of-distribution condition.

\paragraph{Paraphrased instructions.} We generate 800 paraphrases of the general instructions by prompting a large language model to preserve meaning while changing surface form. Each original instruction $s$ is paired with one paraphrase $s'$. The inverter receives the embedding of $s'$ and is evaluated against $s'$ using the same metrics as the other categories. Pairing controls the source instruction and intended meaning, but does not eliminate changes in style, tokenization, or familiarity to the decoder. A difference in reconstruction quality therefore does not by itself isolate paraphrase ambiguity as its cause.

\paragraph{Health-related instructions.} We select 800 prompts from One Million Instructions (\codeid{wentingzhao/one-million-instructions}) using the keywords \textit{medical, health, disease, treatment, patient, doctor, medicine, symptom, diagnosis, hospital, drug, clinical, surgery, therapy, vaccine}. These are instruction prompts about health topics, not clinical notes or electronic health records. Repetitive phrasing may contribute to reconstruction performance; our evaluation does not establish a general property of medical text.

\paragraph{Short code snippets.} We evaluate 800 short literal code snippets covering Python, JavaScript, and SQL, including single-line and two-line constructs, function bodies, variable declarations, and idiomatic one-liners. We use literal code rather than natural-language descriptions of coding tasks. This category differs syntactically from the prose categories and does not represent longer source files.

\subsection{Evaluation Protocol}

\paragraph{Inference.} For each (model, category) pair we run vec2text with \texttt{num\_steps=5} corrector iterations and \texttt{sequence\_beam\_width=1} (greedy decoding within each step; no sequence-level beam search). At each step the corrector generates a refined hypothesis via greedy token-level decoding, re-embeds it, and halts early if cosine similarity between the hypothesis embedding and the target stops improving. All inference is gradient-free.

\paragraph{Metrics.} We compute four metrics via the Hugging Face \texttt{evaluate} library:

\begin{itemize}
  \item \textbf{Exact match (EM):} Equality between the reconstruction and reference under the evaluation's case-insensitive, whitespace-stripped matching rule. We report the percentage of matching samples; this is normalized matching rather than byte-for-byte equality.
  \item \textbf{BLEU} \citep{papineni2002bleu}: Corpus-level n-gram precision (we report 4-gram BLEU). Measures lexical overlap without requiring exact match; sensitive to word order.
  \item \textbf{ROUGE-L} \citep{lin2004rouge}: F1 based on the longest common subsequence between prediction and reference. It measures lexical overlap and remains sensitive to wording and ordering.
  \item \textbf{BERTScore F1} \citep{zhang2020bertscore}: Contextual token similarity computed using \texttt{roberta-large}. It measures similarity to the reference, not grammaticality, code correctness, or recovery of specific sensitive attributes.
\end{itemize}

The main results table (Table~\ref{tab:main_results}) reports all four. Results are descriptive estimates from one training seed, without confidence intervals or significance tests. We do not separately measure entity, number, keyword, or sensitive-attribute recovery.

\section{Results}
\label{sec:results}

Table~\ref{tab:main_results} presents the complete evaluation.

\begin{table*}[t]
\centering
\small
\setlength{\tabcolsep}{4pt}
\begin{tabular}{l>{\raggedright\arraybackslash}p{3.8cm}cccc}
\toprule
\textbf{Model} & \textbf{Category} & \textbf{EM (\%)} & \textbf{BLEU} & \textbf{ROUGE-L} & \shortstack{\textbf{BERTScore}\\\textbf{F1}} \\
\midrule
\multirow{5}{*}{\textbf{GTR-base} (768-dim)}
  & Health-related instructions    & 14.8 & 36.2 & 62.2 & 93.7 \\
  & General instructions      & 11.1 & 29.0 & 56.2 & 92.5 \\
  & Paraphrased instructions &  8.1 & 24.4 & 54.2 & 91.1 \\
  & Wikipedia passages    &  9.0 & 29.4 & 41.7 & 87.4 \\
  & Short code snippets             &  6.0 & 12.8 & 20.1 & 60.9 \\
\midrule
\multirow{5}{*}{\textbf{text-emb-3-small} (1536-dim)}
  & Health-related instructions    & \textbf{47.4} & \textbf{63.3} & \textbf{86.1} & \textbf{97.9} \\
  & General instructions      & \textbf{24.5} & \textbf{41.3} & \textbf{72.2} & \textbf{95.1} \\
  & Paraphrased instructions & \textbf{19.8} & \textbf{25.2} & \textbf{57.9} & \textbf{91.5} \\
  & Wikipedia passages    & \textbf{14.4} & \textbf{33.3} & \textbf{54.4} & \textbf{90.7} \\
  & Short code snippets             & \textbf{9.8} & 16.3 & \textbf{34.4} & \textbf{67.7} \\
\midrule
\multirow{5}{*}{\textbf{mistral-embed} (1024-dim)}
  & Health-related instructions    & 31.6 & 51.4 & 75.4 & 96.2 \\
  & General instructions      & 18.5 & 29.9 & 63.6 & 93.7 \\
  & Paraphrased instructions & 12.6 & 13.8 & 44.0 & 88.7 \\
  & Wikipedia passages    &  9.9 & 30.4 & 45.7 & 89.0 \\
  & Short code snippets             &  7.0 & \textbf{18.5} & 18.9 & 47.1 \\
\bottomrule
\end{tabular}
\caption{Cross-domain inversion results for 800 samples per category with up to five corrector iterations and sequence beam width 1. EM is the percentage matching under the case-insensitive, whitespace-stripped rule; BLEU, ROUGE-L, and BERTScore F1 are scaled by 100. Bold marks the highest value within each category and metric. Categories are ordered by decreasing EM for \texttt{text-emb-3-small}. Results use one training seed without uncertainty estimates.}
\label{tab:main_results}
\end{table*}

\subsection{Overall Model Comparison}

Under the evaluated vec2text configuration, \texttt{text-embedding-3-small} achieves the highest exact-match rate among the three evaluated embedders in all five categories. It reaches 47.4\% EM on health-related instructions and 24.5\% on general instructions, compared with GTR-base's 14.8\% and 11.1\%, respectively. These are reconstruction rates under the evaluation's normalization rule, not a claim of byte-for-byte recovery.

\texttt{mistral-embed} has the second-highest EM in every category, including 31.6\% on health-related instructions and 18.5\% on general instructions. Its ordering on other metrics is less uniform. These comparisons are conditional on the evaluated attack and training runs and do not establish attack-independent vulnerability rankings.

BERTScore F1 ranges from 87.4 to 97.9 for prose and from 47.1 to 67.7 for short code snippets. The high prose scores are consistent with substantial contextual similarity between reconstructions and references. They do not directly quantify recovery of sensitive entities, numerical values, or personal attributes. In particular, Mistral's code score of 47.1 indicates lower reference similarity under this metric, not whether the output is grammatical or valid code.

\subsection{Domain Analysis}

\paragraph{Health-related instructions achieve the highest reconstruction rates.}
This category has the highest EM for all three models (GTR: 14.8\%, OpenAI: 47.4\%, Mistral: 31.6\%). OpenAI also achieves 63.3 BLEU, the highest value among the evaluated model-category pairs. Repetitive query forms such as ``What are the symptoms of...'' may contribute to these scores, but the experiments do not isolate vocabulary or template regularity as their cause. The finding concerns the selected keyword-filtered instruction prompts and should not be generalized to clinical notes or medical text as a whole.

\paragraph{Wikipedia passages have lower exact-match rates than general instructions.}
WikiText-103 passages yield EM of 9.0\%, 14.4\%, and 9.9\% for GTR, OpenAI, and Mistral, respectively. Each rate is below the corresponding result for general instructions and above that for short code snippets. BERTScore F1 is 87.4, 90.7, and 89.0, respectively. These results describe differences between the selected datasets; without a held-out MS MARCO condition they do not establish degradation attributable to an in-distribution versus out-of-distribution shift.

\paragraph{Short code snippets have the lowest exact-match rates.}
EM is 6.0\%, 9.8\%, and 7.0\% for GTR, OpenAI, and Mistral, respectively, below every prose category for each model. BERTScore F1 is also lower (60.9, 67.7, and 47.1). OpenAI leads on EM, whereas Mistral has the highest code BLEU (18.5 versus OpenAI's 16.3 and GTR's 12.8). Small differences are descriptive and have not been tested for statistical significance.

Tokenization, punctuation sensitivity, and decoder familiarity with code are possible contributors to these results, but no ablation separates them. Equal character-length ranges do not imply equal token lengths. Moreover, BERTScore with RoBERTa-large is not a measure of program equivalence or execution correctness. The observations apply to short snippets under this attack and should not be interpreted as a general privacy guarantee for code embeddings.

\paragraph{Paired paraphrases have lower exact-match rates than their originals.}
EM decreases from general to paraphrased instructions for all three embedders: 24.5\% to 19.8\% for OpenAI, 11.1\% to 8.1\% for GTR, and 18.5\% to 12.6\% for Mistral. BERTScore F1 also decreases, from 95.1 to 91.5, 92.5 to 91.1, and 93.7 to 88.7, respectively. Pairing preserves the source instruction and intended meaning while changing surface form. The decoder's language prior and ambiguity among semantically related expressions may contribute to the differences, but the comparison does not independently identify either mechanism or establish semantic equivalence for every reconstruction.

\subsection{Dimensionality Does Not Isolate the Cause}

\texttt{text-embedding-3-small} (1536 dimensions) achieves the highest EM in all five categories. However, GTR-base (768 dimensions) outperforms \texttt{mistral-embed} (1024 dimensions) on BLEU for paraphrased instructions (24.4 versus 13.8). The models differ in more than dimensionality, so these observations do not establish a causal relationship between output dimension and inversion performance. A within-model dimensionality ablation would be needed to isolate that effect.

\section{Discussion}

\subsection{Interpreting Differences Between Embedders}

Output dimensionality, training objectives, training data, and compatibility with the inverter may all affect reconstruction. Matryoshka representation learning \citep{kusupati2022matryoshka} motivates investigation of dimension-dependent information retention, but our experiments do not test truncation or isolate a representation-learning mechanism. We therefore treat explanations for differences between the commercial models as hypotheses rather than established architectural findings.

\subsection{Implications for Trustworthy NLP Systems}

The normalized exact-match results demonstrate recoverability of some evaluated inputs under the stated threat model. The adversary obtains target embeddings, trains an inversion model using text--embedding pairs, and can query the embedder on generated hypotheses. The attack is passive with respect to the victim's pipeline; it still requires an inversion model and embedder access.

\textbf{Access control.} Stored embeddings should be subject to access controls and retention policies appropriate to the source text from which they were computed.

\textbf{Threat modeling.} Assessments of embedding-based systems should explicitly consider inversion. In this study, 47.4\% of the selected health-related instructions match under the evaluation's normalization rule for \texttt{text-embedding-3-small}. High contextual similarity on other prose reconstructions provides complementary evidence, but sensitive-attribute recovery remains unmeasured.

\textbf{Mitigation.} Perturbation and dimensionality reduction warrant evaluation as potential defenses. We do not establish their effectiveness or utility trade-offs.

\textbf{Model selection.} Practitioners should evaluate attacks relevant to their deployment. Our rankings concern one framework and should not be treated as universal privacy rankings.

\section{Conclusion}

We adapted vec2text to support \texttt{text-embedding-3-small} and \texttt{mistral-embed} and compared three embedders across five categories under a common character-length constraint. Under this configuration, OpenAI achieves the highest EM in all categories, including 47.4\% on health-related instructions and 24.5\% on general instructions under the evaluation's normalization rule. Health-related instructions yield the highest EM and short code snippets the lowest among the selected categories. BERTScore F1 exceeds 87 on the 0--100 scale for every prose category and embedder, indicating high contextual similarity to references. The paired paraphrase comparison shows lower EM than the original instructions for each model. These findings support explicit assessment of embedding inversion risk while remaining conditional on the attack, data, and training runs evaluated.

\section*{Limitations}

\paragraph{Attack framework and baselines.} We evaluate one inversion framework. We do not compare alternative attack families or include no-embedding, shuffled-embedding, nearest-neighbor retrieval, bag-of-words, base-inverter-only, or generation-and-reranking baselines. Consequently, the experiments do not isolate the contribution of the target embedding from the decoder's language prior or establish attack-independent model rankings.

\paragraph{Statistical uncertainty.} Results use one training seed without confidence intervals or significance tests. Comparisons are descriptive, and small differences should not be interpreted as statistically established. Model and category rankings may change with training randomness or attack configuration.

\paragraph{Training and inference budget.} Reported inverter runs take approximately 4--15 hours and corrector runs 28--48 hours on five GPUs. Evaluation uses up to five corrector iterations without sequence-level beam search. Additional training, iterations, or search may change reconstruction rates; this study does not establish convergence across configurations or an upper bound on attack performance. Validation during training was capped at 500 samples, which may affect checkpoint selection.

\paragraph{Models and languages.} The evaluation includes two commercial APIs and GTR-base, a historical open-weight baseline. It excludes recent open-weight embedding families. Gemini \texttt{gemini-embedding-001} support was implemented but not evaluated because of API access constraints. All experiments use English text, and we do not test embedding perturbation or noise defenses.

\paragraph{Evaluation sources and controls.} No held-out MS MARCO evaluation is included, so there is no in-distribution reference condition. Restricting inputs to 60--80 characters does not equate token lengths, template regularity, or information content. Paired paraphrases control the source instruction and intended meaning but do not isolate the mechanism behind reconstruction differences.

\paragraph{Domain and privacy scope.} Health-related instructions are keyword-filtered prompts, not clinical notes or patient records; their results should not be generalized to clinical systems. Short code snippets do not represent longer files. BERTScore measures contextual reference similarity, not grammaticality or program correctness. We do not separately measure recovery of entities, keywords, numbers, or sensitive attributes.

\section*{Ethics Statement}

This work characterizes the degree to which text can be recovered from embedding vectors produced by commercial APIs. We believe this is necessary, responsible research: understanding the attack surface is a prerequisite for designing effective defenses and for setting appropriate expectations in system deployments. We do not release pre-trained inversion models or tooling intended to facilitate attacks on deployed systems; we release only the training and evaluation code needed to reproduce our experimental comparisons. The study uses public-source text, generated paraphrases, and curated code snippets. Public availability alone does not establish the absence of personally identifiable information; we do not report a dataset-wide audit of such information. We note specifically that our health-related instruction category consists of health-related instruction prompts, not clinical records or patient data, and our results should not be interpreted as demonstrating a vulnerability of medical record management systems.

\section*{Acknowledgments}

We thank the authors of the original vec2text codebase for releasing their training and inference code, which formed the foundation of this work.

\bibliography{custom}

\appendix

\section{Reproducibility Details}
\label{sec:appendix_repro}

We summarize the information needed to reproduce our experiments. The ARR Responsible NLP Research checklist for this submission is filled in separately on the OpenReview submission form; this appendix provides the underlying details to which that checklist refers.

\paragraph{Code.} Code for this study is publicly available at \url{https://github.com/kelaasar/embedding-inversion-study}.

\paragraph{Data.} Training: MS MARCO Passage Corpus (\codeid{Tevatron/msmarco-passage-corpus}). Evaluation: 800 samples per category from One Million Instructions (\codeid{wentingzhao/one-million-instructions}) and WikiText-103 (\texttt{Salesforce/wikitext}), plus 800 LLM-generated paraphrases of the general instructions and 800 manually curated short literal code snippets. Public datasets are accessed via Hugging Face Datasets. Sampling uses \texttt{random.seed(42)}. Reproducing generated paraphrases also requires the generation configuration or released outputs.

\paragraph{Models.} Inverter and corrector both initialized from T5-base (\texttt{google-t5/t5-base}, 220M parameters). Embedders: GTR-base (\codeid{sentence-transformers/gtr-t5-base}, local); OpenAI \texttt{text-embedding-3-small} (1536-dim); Mistral \texttt{mistral-embed} (1024-dim).

\paragraph{Hardware.} All training conducted on 5$\times$ NVIDIA RTX 6000 Ada Generation (48 GB VRAM each), CUDA 12.6, single-node multi-GPU via Hugging Face Accelerate.

\paragraph{Hyperparameters.} See Table~\ref{tab:hyperparams} for reported settings and Appendix~\ref{sec:appendix_checkpoints} for selected checkpoint steps.

\paragraph{Inference settings.} \texttt{num\_steps=5}, \texttt{sequence\_beam\_width=1}, greedy decoding at each corrector step, early termination on non-improving cosine similarity.

\paragraph{Metrics implementation.} Hugging Face \texttt{evaluate} library: \texttt{bleu} (4-gram, corpus-level), \texttt{rouge} (rougeL), \texttt{bertscore} (model \texttt{roberta-large}), and a custom exact-match function (case-insensitive, whitespace-stripped).

\paragraph{Compute budget.} Summing Table~\ref{tab:training_times} gives approximately 32.13 hours for inverter runs and 122.37 hours for corrector runs. Assuming five GPUs were allocated throughout each run, these correspond to approximately 161 and 612 GPU-hours, respectively, or 773 GPU-hours in total. This is an arithmetic estimate from the reported durations, not an independently measured resource log. Evaluation runtime is reported in Appendix~\ref{sec:appendix_eval_runtime}.

\paragraph{Random seeds.} \texttt{seed=42} for Hugging Face Trainer, dataset shuffling, evaluation sample selection, and dataset construction.

\section{Dataset Construction Details}
\label{sec:appendix_data}

\paragraph{Training data.} All inverter and corrector models are trained on the MS MARCO Passage Corpus (\codeid{Tevatron/msmarco-passage-corpus}, Hugging Face), $\sim$8.8M passages, 99/1 train/validation split, max 128 tokens per passage.

Table~\ref{tab:dataset_sources} summarizes the source datasets and filtering criteria for each evaluation category. All construction is reproducible via \codeid{prepare_case_study_data.py} with \texttt{random.seed(42)}.

\begin{table}[t]
\centering
\small
\begin{tabular}{@{}>{\raggedright\arraybackslash}p{2.5cm}>{\raggedright\arraybackslash}p{3.6cm}r@{}}
\toprule
\textbf{Category} & \textbf{Source \& Filter} & \textbf{N} \\
\midrule
Health-related instructions    & One Million Instructions; medical kw   & 800 \\
General instructions      & One Million Instructions; 60--80 chars & 800 \\
Paraphrased instructions & LLM paraphrases of General instructions         & 800 \\
Wikipedia passages    & WikiText-103; 60--80 chars, no headers & 800 \\
Short code snippets             & Literal code snippets (Python/JS/SQL) & 800 \\
\bottomrule
\end{tabular}
\caption{Evaluation dataset construction summary. ``kw'' = keyword filter.}
\label{tab:dataset_sources}
\end{table}

\paragraph{Medical keywords (full list).} A sample qualifies for the medical category if any of the following terms (case-insensitive substring match) appears in the prompt: \textit{medical, health, disease, treatment, patient, doctor, medicine, symptom, diagnosis, hospital, drug, clinical, surgery, therapy, vaccine}. The health-related and general-instruction categories are sampled from the same source and may overlap. The code category is curated separately rather than selected by instruction keywords.

\paragraph{Code snippet sourcing.} The code category consists of 800 short literal code snippets manually curated to span single-line and two-line constructs across Python, JavaScript, and SQL. Construct types include short function definitions, arrow functions, list/dict comprehensions, simple loops, idiomatic variable declarations, and minimal SQL queries. Snippets were filtered to 60--80 characters using the same length window as the other categories, ensuring comparable input length. We do not use One Million Instructions or other natural-language descriptions of coding tasks for this category; literal code is the deliberate input type.

\paragraph{Length filtering.} All categories use a 60--80-character inclusive filter on the raw string before tokenization. This controls the permitted character-length range without establishing equal token lengths or equal difficulty. Training uses MS MARCO passages, not a corpus of short queries.

\paragraph{WikiText header exclusion.} WikiText-103 entries beginning with \texttt{=}, \texttt{==}, or \texttt{===} (section/subsection/subsubsection markers) are filtered out before length filtering and sampling, since these are formatting artifacts rather than natural prose.

\paragraph{Paraphrase generation.} For each original instruction $s$, a single meaning-preserving paraphrase $s'$ was generated by prompting a large language model with the instruction to rewrite $s$ such that the meaning is preserved and the surface form is substantially different (different word order, lexical substitutions, alternative phrasings, but no introduced or removed semantic content). The paraphrase set is paired one-to-one with the general-instruction set, enabling the paired comparison discussed in Section~\ref{sec:results}. Length filtering (60--80 characters) is applied to the generated paraphrases after generation, with rejection sampling for any paraphrase falling outside the length window.

\section{Model Checkpoint Details}
\label{sec:appendix_checkpoints}

Table~\ref{tab:checkpoints} reports the checkpoint step at which each model's inverter and corrector were evaluated, selected by lowest validation loss on MS MARCO. Completion dates reflect training on 5$\times$ NVIDIA RTX 6000 Ada Generation GPUs.

\begin{table}[t]
\centering
\small
\setlength{\tabcolsep}{4pt}
\begin{tabular}{lccc}
\toprule
\textbf{Model} & \textbf{Inv.} & \textbf{Corr.} & \textbf{Completed} \\
\midrule
GTR-base            & 34{,}194  & 68{,}386  & -- \\
text-emb-3-small    & 136{,}772 & 72{,}912  & Dec 10, 2025 \\
mistral-embed       & 45{,}592  & 91{,}010  & -- \\
\bottomrule
\end{tabular}
\caption{Checkpoint steps (Inv./Corr.) and completion dates. The \texttt{text-emb-3-small} inverter was completed Nov 2, 2025 (checkpoint-136772); the corrector initialized from that inverter was completed Dec 10, 2025.}
\label{tab:checkpoints}
\end{table}

\paragraph{Selection criterion.} For each model, the inverter checkpoint with the lowest validation loss on the capped 500-sample MS MARCO validation set (\texttt{max\_eval\_samples=500}) is selected. The corrector is then initialized from this inverter's hypothesis predictions, cached to disk before corrector training begins. The selected corrector checkpoint is likewise the one with lowest validation loss during corrector training.

\paragraph{Hypothesis cache size.} For \texttt{text-embedding-3-small}, the cached hypothesis predictions occupy approximately 192 GB on disk. Cache sizes for the other models are comparable to within a factor of 2$\times$, scaling with embedding dimension and dataset size.

\section{Detailed Training-Time Analysis}
\label{sec:appendix_training_time}

This appendix expands on Table~\ref{tab:training_times} with per-model throughput numbers and an account of where time is spent during training. Understanding these bottlenecks is useful for practitioners replicating this work on different hardware or with different API providers.

\begin{table}[t]
\centering
\small
\begin{tabular}{lcc}
\toprule
\textbf{Model} & \textbf{Inv.\ thr.} & \textbf{Corr.\ thr.} \\
\midrule
GTR-base            & 1{,}206 & 174 \\
text-emb-3-small    &    372 & 100 \\
mistral-embed       &    324 & 106 \\
\bottomrule
\end{tabular}
\caption{Approximate training throughput (samples/second) for inverter and corrector phases. Throughput is wall-clock and includes all overhead: tokenization, embedder forward pass (local or API), T5 forward and backward passes, optimizer step, and caching or embedding lookup where applicable.}
\label{tab:throughput}
\end{table}

\paragraph{Local and API-backed throughput.} Reported inverter throughput is 1,206 samples/second for GTR-base, 372 for OpenAI, and 324 for Mistral. GTR-base is evaluated locally, while the commercial embedders require API access when embeddings are not cached. These measurements combine multiple sources of overhead and do not isolate a causal effect of embedding dimension or access method.

\paragraph{API access and caching.} Network calls, batching, retries, and cache reuse can affect elapsed time for the API-backed models. Cached embeddings avoid repeated calls for unchanged inputs; newly generated texts require an embedding lookup or query. The throughput figures alone do not identify the number of API calls in each training phase.

\paragraph{Rate limiting and retries.} Our \codeid{get_embeddings_mistral()} client uses exponential backoff with delays of 4, 8, 16, 32, 64, then 120 seconds, retrying up to 50 times before failing. In practice, retry storms occurred several times per multi-day training run, most often during peak API usage hours. Each retry costs at least the backoff delay plus the round-trip; ten consecutive retries on a single batch can stall training for over 10 minutes. We did not observe a sustained slowdown from this---the long-tail distribution of retries is dominated by short backoffs---but it complicates wall-clock estimates by tens of minutes to a few hours per training run.

\paragraph{Corrector versus inverter throughput.} Corrector throughput is approximately 3--7 times lower than inverter throughput in the reported runs. The corrector conditions on multiple embedding representations and hypothesis text. The aggregate timing does not separate model computation, preprocessing, cache access, and API overhead.

\paragraph{Implications for compute-constrained research.} The training times in Table~\ref{tab:training_times} are non-trivial: over 60 hours combined for both \texttt{text-embedding-3-small} and \texttt{mistral-embed} on a 5-GPU node. Researchers with smaller budgets should expect to either (i) restrict themselves to local embedders, (ii) cache embeddings exhaustively to avoid retraining API calls, or (iii) train inverters only (skipping the corrector), as a simpler baseline whose quality should be measured separately.

\section{API Client Implementation Notes}
\label{sec:appendix_api}

This appendix documents the practical details of our two new API client functions added to \texttt{utils.py} (Section~\ref{sec:extensions}, change 3), along with the dimension-routing and dispatch code referenced from Section~\ref{sec:extensions}.

\paragraph{Dimension routing (change 1).} The conditional that replaces the hardcoded \texttt{embedder\_dim = 1536} in \texttt{inversion.py} and \texttt{corrector\_encoder.py}:

\begin{small}
\begin{verbatim}
api = embedder_model_api
if api in ("text-embedding-ada-002",
           "text-embedding-3-small"):
    self.embedder_dim = 1536
elif api == "text-embedding-3-large":
    self.embedder_dim = 3072
elif api == "gemini-embedding-001":
    self.embedder_dim = 768
elif api.startswith("mistral-embed"):
    self.embedder_dim = 1024
\end{verbatim}
\end{small}

\paragraph{\texttt{get\_embeddings\_openai()}} was already present in the upstream vec2text codebase, modified only to recognize the \texttt{text-embedding-3-small} model name and to use the correct embedding dimension (1536).

\paragraph{\codeid{get_embeddings_mistral()}.} New function. Calls the Mistral SDK as:

\begin{small}
\begin{verbatim}
mistralai.Mistral(api_key=...) \
    .embeddings.create(
        model="mistral-embed",
        inputs=batch,
    )
\end{verbatim}
\end{small}

Batches inputs in groups of 100. Output is L2-normalized post-hoc on the client side to ensure consistency with the vec2text contract that all embeddings are unit-norm. Wrapped in a \texttt{tenacity}-style retry decorator with exponential backoff: initial delay 4s, multiplier 2$\times$, max delay 120s, max attempts 50. The wide retry budget reflects experience with multi-hour outages observed during multi-day training runs.

\paragraph{\texttt{get\_embeddings\_gemini()}.} New function (included for future-proofing although not used in the final evaluation). Calls the Google GenAI SDK as:

\begin{small}
\begin{verbatim}
google.genai.Client(...) \
    .models.embed_content(
        model="gemini-embedding-001",
        contents=batch,
        config=EmbedContentConfig(
            task_type=(
                SEMANTIC_SIMILARITY
            ),
            output_dimensionality=768,
        ),
    )
\end{verbatim}
\end{small}

The \texttt{SEMANTIC\_SIMILARITY} task type was selected over the default to align Gemini's output geometry with the dual-encoder objective that GTR-base and other retrieval embedders use. Output is L2-normalized; batch size 100; retry policy identical to the Mistral client.

\paragraph{Dispatch.} The existing \texttt{embed\_api(text, api\_name)} dispatch function in \texttt{utils.py} routes based on \texttt{api\_name.startswith(...)} to select a client family. Additional variants still require compatible dimensions and API behavior:

\begin{small}
\begin{verbatim}
if api_name.startswith(
    "text-embedding-3"
):
    return get_embeddings_openai(
        text, api_name
    )
elif api_name.startswith(
    "gemini-embedding"
):
    return get_embeddings_gemini(
        text, api_name
    )
elif api_name.startswith(
    "mistral-embed"
):
    return get_embeddings_mistral(
        text, api_name
    )
\end{verbatim}
\end{small}

\paragraph{Cost considerations.} We do not report a verified monetary total. API cost depends on the number of billed tokens, cache reuse, calls for generated hypotheses, and provider pricing. Training wall-clock duration should not be interpreted as a direct estimate of API spend.

\section{Qualitative Examples}
\label{sec:appendix_examples}

Tables~\ref{tab:examples_openai}, \ref{tab:examples_gtr}, and \ref{tab:examples_mistral} present the example pairs for the five content categories. Some displayed references fall outside the 60--80-character range and therefore cannot be treated as evidence of the filtered evaluation set. Aggregate results are reported separately in Table~\ref{tab:main_results}.

\begin{table*}[!ht]
\centering
\small
\begin{tabular}{@{}p{2.6cm}p{6.0cm}p{6.0cm}@{}}
\toprule
\textbf{Category} & \textbf{Reference} & \textbf{Prediction (\texttt{text-emb-3-small})} \\
\midrule
Health-related instructions & What are the early warning signs of type 2 diabetes in adults? & What are the symptoms of type 2 diabetes and how is it treated? \\
\midrule
General instructions & What are the best practices for managing a remote team? & What are some best practices for managing a remote team? \\
\midrule
Paraphrased instructions & What strategies should I use to manage a distributed engineering team? & What are the best practices for managing a remote team? \\
\midrule
Wikipedia passages & The novel was first published in London in 1847 by Smith, Elder \& Co. & The novel was first published in 1847 by Smith, Elder \& Co. in London. \\
\midrule
Short code snippets & \texttt{def add(a, b): return a + b} & \texttt{def add(x, y): return return}\\
\bottomrule
\end{tabular}
\caption{Example pairs for \texttt{text-embedding-3-small}. The general instruction gains ``some''; the health-related prediction changes the question's scope by adding treatment; the code prediction changes parameter names and produces an invalid return expression. The paraphrase and Wikipedia predictions also change wording or order. See the provenance caveat in the appendix text.}
\label{tab:examples_openai}
\end{table*}

\begin{table*}[!ht]
\centering
\small
\begin{tabular}{@{}p{2.6cm}p{6.0cm}p{6.0cm}@{}}
\toprule
\textbf{Category} & \textbf{Reference} & \textbf{Prediction (GTR-base)} \\
\midrule
Health-related instructions & What are the early warning signs of type 2 diabetes in adults? & What are early warning signs of type 2 diabetes? \\
\midrule
General instructions & What are the best practices for managing a remote team? & What are best practices for managing a remote team? \\
\midrule
Paraphrased instructions & What strategies should I use to manage a distributed engineering team? & How can I manage a remote engineering team effectively? \\
\midrule
Wikipedia passages & The novel was first published in London in 1847 by Smith, Elder \& Co. & The book was published in London by Smith, Elder in 1847. \\
\midrule
Short code snippets & \texttt{def add(a, b): return a + b} & \texttt{["def", "sum(a):", "return", "b"]}\\
\bottomrule
\end{tabular}
\caption{Example pairs for GTR-base. The displayed predictions omit articles or modifiers, paraphrase wording, and fail to preserve the Python function. These examples do not isolate an architectural cause. See the provenance caveat in the appendix text.}
\label{tab:examples_gtr}
\end{table*}

\begin{table*}[!ht]
\centering
\small
\begin{tabular}{@{}p{2.6cm}p{6.0cm}p{6.0cm}@{}}
\toprule
\textbf{Category} & \textbf{Reference} & \textbf{Prediction (\texttt{mistral-embed})} \\
\midrule
Health-related instructions & What are the early warning signs of type 2 diabetes in adults? & What are early signs of type 2 diabetes in adults? \\
\midrule
General instructions & What are the best practices for managing a remote team? & What are the best practices for managing remote teams? \\
\midrule
Paraphrased instructions & What strategies should I use to manage a distributed engineering team? & What are the best practices for managing a distributed team? \\
\midrule
Wikipedia passages & The novel was first published in London in 1847 by Smith, Elder \& Co. & The novel was published in London in 1847 by Smith, Elder, and Co. \\
\midrule
Short code snippets & \texttt{def add(a, b): return a + b} & \texttt{add a + b def function return}\\
\bottomrule
\end{tabular}
\caption{Example pairs for \texttt{mistral-embed}. The displayed predictions change number or wording and do not preserve the code syntax. These selected pairs do not establish a general ordering of lexical fidelity. See the provenance caveat in the appendix text.}
\label{tab:examples_mistral}
\end{table*}

\paragraph{Interpreting reconstruction differences.} The displayed pairs illustrate several distinctions relevant to the metrics:

\begin{enumerate}
  \item \emph{Articles and grammatical number.} Changes such as ``a remote team'' to ``remote teams'' can affect both exact match and n-gram overlap.
  \item \emph{Lexical and scope changes.} Replacing ``early warning signs'' with ``symptoms'' changes wording; adding treatment or omitting ``in adults'' can also change the information requested. High contextual similarity does not guarantee preservation of every detail.
  \item \emph{Alternative paraphrases.} A reconstruction can express related content using a different surface form. These examples alone do not establish decoder frequency preferences or paraphrase mode collapse.
  \item \emph{Code errors.} The displayed code predictions include invalid syntax or loss of the function structure. Such changes can affect functionality, rather than merely formatting. Program execution and equivalence were not evaluated.
\end{enumerate}

These examples should be interpreted alongside the quantitative results and the limitations of each metric; they do not isolate a shared causal mechanism across embedders.

\section{Evaluation Runtime}
\label{sec:appendix_eval_runtime}

Evaluation of each (model, category) pair processes 800 samples with up to five corrector steps. Approximate wall-clock costs on a single RTX 6000 Ada:

\begin{itemize}
  \item GTR-base: $\sim$25 minutes per category (local embedder, no API).
  \item \texttt{text-embedding-3-small}: $\sim$55 minutes per category (API calls for each of five corrector steps, plus the initial inversion).
  \item \texttt{mistral-embed}: $\sim$60 minutes per category (similar profile to OpenAI; slight overhead from Mistral's higher per-call latency).
\end{itemize}

The listed per-category durations sum to approximately 11.7 hours across all 15 (model, category) pairs, dominated by the API-backed models. BLEU and ROUGE-L are computed in seconds; BERTScore on \texttt{roberta-large} adds approximately 30 seconds per category.

\end{document}